\documentclass[a4paper,10pt]{amsart}
\usepackage[ansinew]{inputenc}
\usepackage{graphicx,psfrag}
\usepackage{xspace}
\usepackage{booktabs}
\usepackage{amssymb}
\usepackage{amsthm}
\usepackage{textcomp}
\usepackage{verbatim}

\newtheorem{thm}{Theorem}[section]

\newtheorem{lem}[thm]{Lemma}
\newtheorem{prop}[thm]{Proposition}
\newtheorem{ob}[thm]{Observation}
\theoremstyle{definition}
\newtheorem*{defn}{Definition}
\newtheorem*{rem}{Remark}

\theoremstyle{remark}
\newtheorem*{claim}{Claim}

\begin{document}

\newcommand{\set}[1]{\left\{#1\right\}}
\newcommand{\maj}[1]{\overline{#1}}
\renewcommand{\min}[1]{\underline{#1}}
\newcommand{\sub}[1]{\ensuremath{\mathrm{#1}}}

\newcommand{\size}[1]{\ensuremath{\lVert#1\rVert}}
\newcommand{\odd}[2]{\ensuremath{[#1,#2]_{\sf{odd}}}}
\newcommand{\even}[2]{\ensuremath{[#1,#2]_{\sf{even}}}}
\newcommand{\interval}[2]{\ensuremath{[#1,#2]}}
\newcommand{\Rodd}[2]{\ensuremath{R(\odd{#1}{#2})}}
\newcommand{\Godd}[2]{\ensuremath{G(\odd{#1}{#2})}}
\newcommand{\Reven}[2]{\ensuremath{R(\even{#1}{#2})}}
\newcommand{\Geven}[2]{\ensuremath{G(\even{#1}{#2})}}

\newcommand{\Lk}{\ensuremath{C_{{\rm{mid}}}}\xspace}
\newcommand{\LL}{\ensuremath{\mathcal{L}}\xspace}

\newcommand{\Lbest}{\ensuremath{C_{{\rm{best}}}}\xspace}
\newcommand{\Lworst}{\ensuremath{C_{{\rm{worst}}}}\xspace}
\newcommand{\LBest}{\ensuremath{\mathcal{C}_{{\rm{best}}}}\xspace}
\newcommand{\LWorst}{\ensuremath{\mathcal{C}_{{\rm{worst}}}}\xspace}

\newcommand{\fb}{{\rm{fB}}\xspace}
\newcommand{\mfb}[1]{\ensuremath{{\rm{mfB}_{#1}}}\xspace}

\renewcommand{\leq}{\leqslant}
\renewcommand{\geq}{\geqslant}

\title{How to eat $4/9$ of a pizza}
\author{Kolja Knauer \and Piotr Micek \and Torsten Ueckerdt}

\begin{abstract}
Two players want to eat a sliced pizza by alternately picking its pieces. The pieces may be of various sizes. After the first piece is eaten every subsequently picked piece must be adjacent to some previously eaten. We provide a strategy for the starting player to eat $\frac{4}{9}$ of the total size of the pizza. This is best possible and settles a conjecture of Peter Winkler.
\end{abstract}

\maketitle

\section{The Problem}

\noindent Alice and Bob share a pizza. The pizza is sliced by 
cuts from the middle to the crust. There may be
any number of pieces which may be of various sizes. To eat the pizza Alice and
Bob have to stick to the following politeness protocol:
\begin{enumerate}
 \item They pick pieces in an alternating fashion;
 \item Alice starts by eating any piece of the pizza;
 \item Afterwards only pieces adjacent to already-eaten pieces may be picked.
\end{enumerate}
This means that on each turn (except the first and the last) a player has two available pieces from which to pick.

\begin{figure}[ht]
 \psfrag{A1}[cc][cc]{Alice's 1st}
 \psfrag{A2}[cr][cr]{Alice's 2nd}
 \psfrag{A3}[cr][cr]{Alice's 3rd}
 \psfrag{B1}[cl][cl]{Bob's 1st}
 \psfrag{B2}[cr][cr]{Bob's 2nd}
 \psfrag{B3}[cl][cl]{Bob's 3rd}
 \centering
 \includegraphics[width = .7\textwidth]{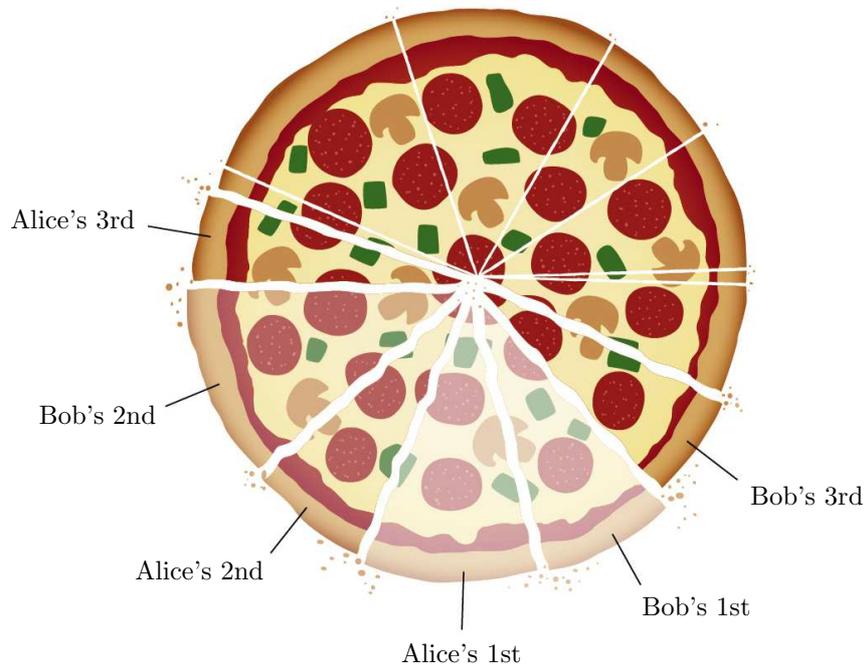}
 \caption{An example showing the first steps in the pizza game.}
\label{fig:first}
\end{figure}

This paper deals with the following question: How should Alice pick her pieces to eat 
a big portion of the pizza? We develop a strategy for her that guarantees her at least $\frac{4}{9}$ of the whole pizza. The strategy works for every possible cutting of the pizza and for every possible behavior of Bob. The ratio $\frac{4}{9}$ is best possible; examples where Alice cannot eat more of the pizza were previously known \cite{Win}.

A peculiarity of our pizzas is that they are allowed to have pieces of zero size. If one prefers, such pieces can be thought of as having very small $\varepsilon$-size, though the importance of such pieces is to the structure, not to the size, of the pizza.\footnote{We discuss this issue in more detail in the final remarks at the end of the paper.} Generally, for a set $S$ of pieces we refer to the sum of sizes of its elements as its \emph{size} $\size{S}$. If we consider the number of pieces in such a set, we make that clear. We are only interested in the portion of the pizza that Alice can eat and hence we assume w.l.o.g. that the size of the whole pizza is $1$.

A simple and nice argument yields the following.
\begin{prop}\label{prop:even-pizzas}
Alice can eat at least $\frac{1}{2}$ of a pizza with an even number of pieces.
\end{prop}
\begin{proof}
Color the pieces alternately green and red. This is possible as the number of pieces is even. Either the green or the red pieces carry at least $\frac{1}{2}$, say the red part. To eat all the red pieces, Alice starts with any red piece and then she always picks the piece which was just revealed by Bob. In this way Alice leaves only green pieces for Bob.
\end{proof}

At first glance the case of pizzas with an odd number of pieces looks better for Alice. She eats one piece more than Bob. Curiously, things can get worse for her (see Proposition~\ref{prop:4-9-example}). The rest of the paper will deal exclusively with pizzas with an odd number of pieces. In Proposition~\ref{prop:1-3} we show that the argument applied for pizzas with an even number of pieces can be adapted to guarantee $\frac{1}{3}$ of the pizza for Alice in the odd case. To this end we introduce some notation.

\begin{figure}[ht]
 \psfrag{C1}[cc][cc]{$C_1$}
 \psfrag{C2}[cc][cc]{$C_2$}
 \psfrag{C3}[cc][cc]{$C_3$}
 \psfrag{C4}[cc][cc]{$C_4$}
 \psfrag{C}[cc][cc]{$C$}
 \psfrag{CC}[cc][cc]{$\odd{C}{C}$}
 \psfrag{C1C2}[cc][cc]{$\odd{C_1}{C_2}$}
 \psfrag{C3C4}[cc][cc]{$\even{C_3}{C_4}$}
 \centering
 \includegraphics[width=0.8\textwidth]{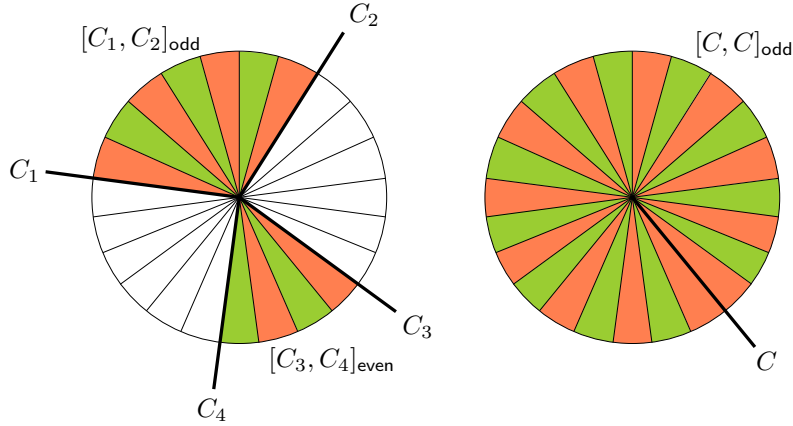}
 \caption{Three intervals and their canonical colorings.}
 \label{fig:intervals}
\end{figure}

By an \emph{interval} of a pizza we mean a set of consecutive pieces. Odd and even intervals are those with an odd and an even number of pieces, respectively. Any interval is bounded by two cuts. Since the pizza has an odd number of pieces any two cuts $C_1$ and $C_2$ enclose one odd and one even interval which we denote by $\odd{C_1}{C_2}$ and $\even{C_1}{C_2}$, respectively. We consider every interval with a \emph{canonical coloring} of its pieces as follows. The pieces of $\odd{C_1}{C_2}$ and $\even{C_1}{C_2}$ are alternately colored red and green starting with a red piece adjacent to $C_1$. Note that for an even interval the order of its bordering cuts is crucial as the red pieces of $\even{C}{C'}$ are the green pieces of $\even{C'}{C}$ and vice versa. Two intervals and their canonical colorings are illustrated\footnote{For a better accessibility of our figures all pieces are drawn equally sized. When needed, we refer to the size of a piece by putting a non-negative number into it.} in Figure~\ref{fig:intervals}. We denote the set of red and green pieces of odd and even intervals by $\Rodd{C_1}{C_2}$ and $\Godd{C_1}{C_2}$, and $\Reven{C_1}{C_2}$ and $\Geven{C_1}{C_2}$, respectively.

Now since the pizza has an odd number of pieces it can be seen as an odd interval on its own. Indeed, there are several odd intervals representing the whole pizza and every such is of the form $\odd{C}{C}$, where $C$ is just a single cut. The key insight is that Alice can force the game to end up with Alice's and Bob's pieces being $\Rodd{C}{C}$ and $\Godd{C}{C}$, respectively, for some cut $C$. She can do so by behaving like in the previous proof: after the first piece Alice always picks the piece which was just revealed by Bob. Such a strategy for Alice is called \emph{follow-Bob}, shortly \fb.



\begin{prop}\label{prop:1-3}
 Alice can eat at least $\frac{1}{3}$ of a pizza with an odd number of pieces.
\end{prop}
\begin{proof}
 Choose a cut $C$ such that $\size{\Rodd{C}{C}}$ is minimal. By playing any \fb-strategy Alice eats at least $\size{\Rodd{C}{C}}$, so assume $\size{\Rodd{C}{C}} < \frac{1}{3}$ and hence $\size{\Godd{C}{C}} > \frac{2}{3}$. Let $p$ be a green piece such that the size of the green pieces from $p$ (included) to the cut $C$ in either direction is at least $\frac{1}{2}\size{\Godd{C}{C}}$. Now let Alice start with $p$ and play \fb. This way Alice eats all green pieces from $p$ to $C$ in at least one direction and so she eats at least $\frac{1}{2}\cdot\frac{2}{3}$ of the pizza.
\end{proof}

\begin{figure}[ht]
 \psfrag{1}[cc][cc]{$1$}
 \psfrag{0}[cc][cc]{$0$}
 \centering
 \includegraphics[width = .32\textwidth]{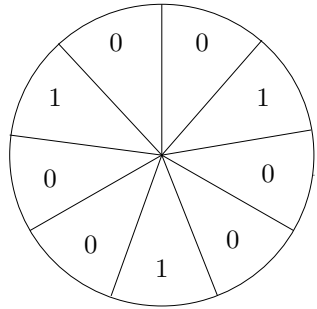}
 \caption{A pizza in which Alice eats only $\frac{1}{3}$ playing \fb. The numbers stand for piece sizes.}
 \label{fig:follow-Bob-killer}
\end{figure}

Proposition~\ref{prop:1-3} shows that there always exists an \fb-strategy that enables Alice to eat at least $\frac{1}{3}$ of the pizza. This is the best Alice can ensure by playing \fb. To see this consider the pizza depicted in Figure~\ref{fig:follow-Bob-killer} that allows Bob to always eat $\frac{2}{3}$ of the pizza if Alice plays \fb. On the other hand it is easy to see that Alice can prevent Bob from eating more than $\frac{1}{3}$ of this particular pizza, but to this end Alice has to come up with a different strategy from simply following Bob.

\begin{figure}[ht]
 \psfrag{2}[cc][cc]{$2$}
 \psfrag{1}[cc][cc]{$1$}
 \psfrag{0}[cc][cc]{$0$}
 \psfrag{L_1}[cc][cc]{$ $}
 \psfrag{L_k}[cc][cc]{$ $}
 \psfrag{L_l}[cc][cc]{$ $}
 \centering
 \includegraphics[width = .49\textwidth]{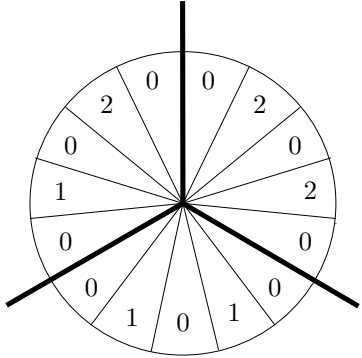}
 \caption{A pizza of which every strategy ensures at most $\frac{4}{9}$ for Alice.}
 \label{fig:4-9-example}
\end{figure}

Unfortunately there are also pizzas in which, if Bob is very smart, Alice cannot eat half of the total size. The example presented in Figure~\ref{fig:4-9-example} is due to Peter Winkler. In fact, there is even a $\set{0,1}$-pizza (with pieces of sizes $0$ and $1$) with $21$ pieces of which Alice eats at most $\frac{4}{9}$ against a clever Bob. The upcoming methods in this paper can be used to show the minimality of these examples in terms of number of pieces. Finally note that, in general, Alice can find an optimal strategy for a fixed pizza by a dynamic programming approach in quadratic time.

\begin{prop}\label{prop:4-9-example}
 There are pizzas of which Bob can eat $\frac{5}{9}$.
\end{prop}

\begin{proof}
 Consider the pizza from Figure \ref{fig:4-9-example}. The size of the pizza is $9$, so we provide a strategy for Bob to eat pieces whose sizes sum up to at least $5$.

 If Alice starts with a $0$-piece, then the remaining part has an even number of pieces and still has size $9$. So Bob can two-color the pieces and eat the color with larger size as Alice did in Proposition~\ref{prop:even-pizzas}. In this way Bob's outcome is at least $\lceil\frac{9}{2}\rceil=5$.

 In order to deal with a different behavior of Alice, consider the partition of the pizza into the three odd intervals indicated by the three thick cuts in Figure~\ref{fig:4-9-example}. If Alice starts with a non-zero piece, Bob picks the available piece adjacent to a thick cut. Afterwards Bob always picks the piece just revealed by Alice (so he follows Alice) unless this would mean eating from a still untouched interval. If both pieces available to Bob are from untouched intervals, he picks the piece from the interval of smaller size. One can verify (several elementary cases) that Bob always eats at least $5$ with this strategy.
\end{proof}


At \textquotedblleft Building Bridges: a conference on mathematics and computer science in honour of Laci Lov\'{a}sz\textquotedblright, in Budapest, August 5-9 2008, Peter Winkler conjectured that \emph{Alice can eat at least $\frac{4}{9}$ of any pizza}. He also noted that $\frac{1}{3}$ from below is easy and $\frac{4}{9}$ is best possible. We verify the conjecture to be true. Independently, the same result is given by Josef Cibulka, Jan Kyn\v{c}l, Viola M\'{e}sz\'{a}ros, Rudolf Stola\'{r} and Pavel Valtr~\cite{CKMSV}.

We already pointed out that in order to eat more than $\frac{1}{3}$ of the pizza Alice has to find strategies different from \fb. Nevertheless the best \fb-strategy can be really valuable to Alice. Our arguments for strategies better than $\frac{1}{3}$ consider several strategies, at least one of which turns out to be good depending on the pizza. A certain \fb-strategy will always be a candidate.

Based on a strong connection between \fb-strategies and odd intervals, Section~\ref{sec:SLA} focuses on how the structure of the pizza can be analyzed relative to its odd intervals. We will show that either a pizza is \emph{easy} for Alice or we can partition it into three nicely structured odd intervals that form the foundation of all our strategies. In Section~\ref{sec:three} we slightly modify the \fb-strategies based on the above-mentioned intervals. We will prove that the best of \fb-strategies and modified-\fb-strategies yields at least $\frac{3}{7}$ of the pizza for Alice. Finally, in Section~\ref{sec:four} we refine the idea underlying the modified-\fb-strategies. This results in a new set of strategies and the outcome of $\frac{4}{9}$ of the pizza for Alice.

\section{Partitioning the pizza}\label{sec:SLA}

\noindent Remember that for any cut $C$ we may consider the pizza as the odd interval $\odd{C}{C}$ with its canonical coloring into red and green pieces (see the right of Figure~\ref{fig:intervals}). If Alice plays \fb the resulting distribution is $\Rodd{C}{C}, \Godd{C}{C}$ for some cut $C$, no matter what Bob does, where Alice and Bob eat red and green, respectively.

\begin{figure}[ht]
 \psfrag{C}[cc][cc]{$C$}
 \psfrag{C1}[cc][cc]{$C_1$}
 \psfrag{C2}[cc][cc]{$C_2$}
 \psfrag{C1C2}[cc][cc]{$\odd{C_1}{C_2}$}
 \psfrag{CC1}[cc][cc]{$\even{C}{C_1}$}
 \psfrag{CC2}[cc][cc]{$\even{C}{C_2}$}
 \centering
 \includegraphics[width=0.46\textwidth]{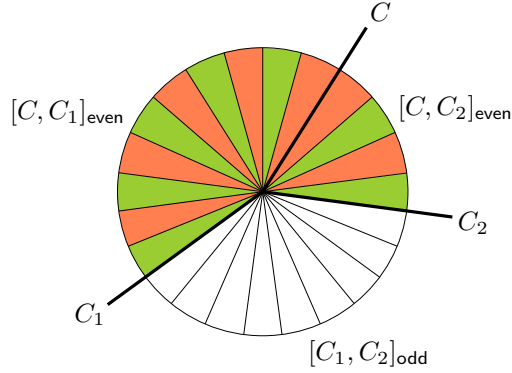}
 \caption{Alice follows Bob after $\odd{C_1}{C_2}$ is eaten. This results in a cut $C$ spliting $\even{C_1}{C_2}$ into two even intervals. Alice gets the red pieces $\Reven{C}{C_1} \cup \Reven{C}{C_2}$ and Bob gets the green pieces $\Geven{C}{C_1} \cup \Geven{C}{C_2}$.}
 \label{fig:best-answer}
\end{figure}

Let us slightly generalize this. Consider an intermediate point in the game in which it is Bob's turn, i.e., the pieces of an odd interval $\odd{C_1}{C_2}$ are already eaten. We say that Alice \emph{follows Bob after $\odd{C_1}{C_2}$ is eaten}, if in every further turn she picks the piece that was just revealed by Bob. As a consequence the set of \emph{remaining} pieces, namely $\even{C_1}{C_2}$, will be distributed among Alice and Bob in the following fashion (see Figure~\ref{fig:best-answer} for an example).
There will be a cut $C$ splitting $\even{C_1}{C_2}$ into two even intervals, whereas Alice gets $\Reven{C}{C_1} \cup \Reven{C}{C_2}$ and Bob gets $\Geven{C}{C_1} \cup \Geven{C}{C_2}$. With this terminology an \fb-strategy with starting piece $p$ means that Alice follows Bob after $\{p\} = \odd{C_1}{C_2}$ is eaten.

Now, for ending up with the cut $C$ there are many possible behaviors of Bob but all of them yield the same distribution of Alice's and Bob's pieces. We are not interested in the exact \textit{course} of an \fb-strategy, but in the outcome in terms of the resulting canonical colorings.

\begin{ob}
 When Alice follows Bob after $\odd{C_1}{C_2}$ is eaten, then Bob's behaviour can be reduced to the choice of a cut $C$ splitting $\even{C_1}{C_2}$ into two even intervals: $\even{C}{C_1}$ and $\even{C}{C_2}$. 
Then of the remaining pieces Alice gets $\Reven{C}{C_1} \cup \Reven{C}{C_2}$ and Bob gets $\Geven{C}{C_1} \cup \Geven{C}{C_2}$.
\end{ob}


For any given odd interval $\odd{C_1}{C_2}$ of already eaten pieces there are cuts $C$, which minimize $\size{\Reven{C}{C_1} \cup \Reven{C}{C_2}}$, namely Alice's outcome, among all cuts $C$ with $\even{C}{C_1} \subseteq \even{C_1}{C_2}$. We call such a cut a (Bob's) \emph{best answer} to $\odd{C_1}{C_2}$. A given odd interval may have several best answers and a single cut may be a best answer to several intervals. Most importantly, best answers can be characterized using the following definition.

\begin{defn}
 An even interval $\even{C_1}{C_2}$ has the \emph{heavy greens property} if for every $\even{C_1}{C} \subseteq \even{C_1}{C_2}$ we have $$\size{\Geven{C_1}{C}} \geq \size{\Reven{C_1}{C}}.$$
 An odd interval $\odd{C_1}{C_2}$ has the \emph{heavy greens property} if additionally for every $\even{C_2}{C} \subseteq \odd{C_1}{C_2}$ we have $$\size{\Geven{C_2}{C}} \geq \size{\Reven{C_2}{C}}.$$
\end{defn}

Note that in case of an even interval the heavy greens property, just as the canonical coloring, depends on the order of the bordering cuts. I.e., $\even{C_1}{C_2}$ having the heavy greens property is not the same as $\even{C_2}{C_1}$ having it.

\begin{lem}\label{lem:greens}
 A cut $C$ with $\even{C}{C_1} \subseteq \even{C_1}{C_2}$ is a best answer to $\odd{C_1}{C_2}$ if and only if $\even{C}{C_1}$ and $\even{C}{C_2}$ have the heavy greens property.
\end{lem}

\begin{figure}[ht]
 \psfrag{C}[cc][cc]{$C$}
 \psfrag{C1}[cc][cc]{$C_1$}
 \psfrag{C2}[cc][cc]{$C_2$}
 \psfrag{Ct}[cc][cc]{$\tilde{C}$}
 \psfrag{C1C2}[cc][cc]{$\odd{C_1}{C_2}$}
 \psfrag{CtC1}[cc][cc]{$\even{\tilde{C}}{C_1}$}
 \psfrag{CCt}[cc][cc]{$\even{C}{\tilde{C}}$}
 \psfrag{CC2}[cc][cc]{$\even{C}{C_2}$}
 \centering
 \includegraphics[width=0.5\textwidth]{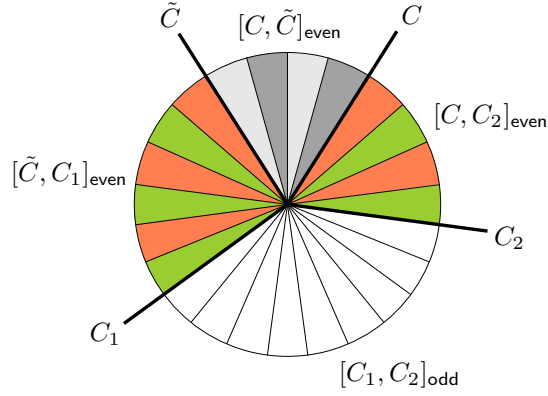}
 \caption{Two cuts $C$ and $\tilde{C}$ which are possible best answers to $\odd{C_1}{C_2}$. The set $\Reven{C}{C_1} \cup \Reven{C}{C_2}$ is the union of red and dark grey pieces, the set $\Reven{\tilde{C}}{C_1} \cup \Reven{\tilde{C}}{C_2}$ is the union of red and light grey pieces.}
 \label{fig:greens}
\end{figure}

\begin{proof}
 By definition a cut $C$ is a best answer to $\odd{C_1}{C_2}$ if and only if it minimizes $\size{\Reven{C}{C_1} \cup \Reven{C}{C_2}}$ among all cuts $C$ with $\even{C}{C_1} \subseteq \even{C_1}{C_2}$. That is for every other cut $\tilde{C}$ with $\even{C}{\tilde{C}} \subseteq \even{C_1}{C_2}$ we have
 \begin{equation}
  \size{\Reven{C}{C_1} \cup \Reven{C}{C_2}} \geq \size{\Reven{\tilde{C}}{C_1} \cup \Reven{\tilde{C}}{C_2}}.\label{eq:greens}
 \end{equation}

 \noindent As illustrated in Figure~\ref{fig:greens} the symmetric difference of Alice's pieces w.r.t. $C$ and $\tilde{C}$ is precisely the set $\even{C}{\tilde{C}}$. In particular $\Geven{C}{\tilde{C}} \subseteq \Reven{C}{C_1} \cup \Reven{C}{C_2}$ and $\Reven{C}{\tilde{C}} \subseteq \Reven{\tilde{C}}{C_1} \cup \Reven{\tilde{C}}{C_2}$. Thus~\eqref{eq:greens} is equivalent to
 $$\size{\Geven{C}{\tilde{C}}} \geq \size{\Reven{C}{\tilde{C}}},$$
 which is the heavy greens property.
\end{proof}


If the game comes to a point at which precisely $\odd{C_1}{C_2}$ is eaten and $C$ is a best answer to $\odd{C_1}{C_2}$, then Alice can follow Bob from then on and thus guarantee herself at least $\size{\Reven{C}{C_1} \cup \Reven{C}{C_2}}$ within the remaining pieces. In the special case of $\odd{C_1}{C_2}$ being just a single piece $p$ we will refer to this strategy as an \fb-strategy \emph{associated} with the cut $C$.

\begin{defn}
We call a pizza \emph{easy} if there is an \fb-strategy yielding at least $\frac{1}{2}$ of the pizza for Alice. Otherwise we call the pizza \emph{hard}.
\end{defn}

Actually, we have already noted that there are pizzas with no \fb-strategy yielding more than $\frac{1}{3}$ (see Figure~\ref{fig:follow-Bob-killer}). The rest of this section is dedicated to prove the following theorem.

\begin{thm}\label{thm:main}
 A hard pizza can be partitioned into three odd intervals each having the heavy greens property.
\end{thm}

We will need another lemma and two definitions. At first, call two distinct cuts \emph{neighboring} if they enclose a single piece of the pizza.

\begin{lem}\label{lem:adjacent-lines}
 If two neighboring cuts are best answers to a single piece each, then the pizza is easy.
\end{lem}

\begin{figure}[ht]
 \psfrag{p}[cc][cc]{$p$}
 \psfrag{C1}[cc][cc]{$C$}
 \psfrag{C2}[cc][cc]{$C'$}
 \centering
 \includegraphics[width = .35\textwidth]{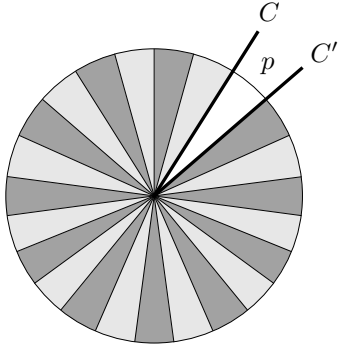}
 \caption{A pizza with neighboring best answers $C$ and $C'$. The set $\Rodd{C}{C}$ consists of $p$ and all light grey pieces, the set $\Rodd{C'}{C'}$ consists of $p$ and all dark grey pieces.}
 \label{fig:neighboring}
\end{figure}

\begin{proof}
Let $p$ be the piece between two neighboring best answers $C$ and $C'$. As $\Rodd{C'}{C'}=\Godd{C}{C}\cup\{p\}$ (see Figure~\ref{fig:neighboring}), we get that $\Rodd{C}{C}\cup \Rodd{C'}{C'}$ is the whole pizza. This implies that the size of one of the two -- say $\size{\Rodd{C}{C}}$ -- is at least $\frac{1}{2}$. But since $C$ is a best answer, Alice playing an \fb-strategy associated to $C$ eats at least $\size{\Rodd{C}{C}}$.
\end{proof}

Consider the set \LWorst of those cuts $C$ which minimize $\size{\Rodd{C}{C}}$ among all cuts. Clearly, a cut $C \in \LWorst$ is a best answers to every piece $p \in \Rodd{C}{C}$.


Since it is needed in Section~\ref{sec:four} we prove a stronger statement than Theorem~\ref{thm:main}, namely that the cuts defining the tripartition can be chosen to be best answers to single pieces, one of them being a cut in $\LWorst$ of our choice.

\begin{thm}\label{thm:main-more}
 For every hard pizza and every $C_1 \in \LWorst$ there are two further best answers $C_2$ and $C_3$, such that $\odd{C_1}{C_2}$, $\odd{C_1}{C_3}$, and $\odd{C_2}{C_3}$ are disjoint and each has the heavy greens property.
\end{thm}

\begin{proof}
 In contrast to $\LWorst$, define $\LBest$ to be the set of all cuts $C$ which \emph{maximize} $\size{\Rodd{C}{C}}$ among all best answers to a single piece. Furthermore let $A(C)$ denote the set of single pieces to which a given cut $C$ is a best answer. Given $C_1 \in \LWorst$, the two further best answers are chosen as follows:

 \begin{list}{\labelitemi}{\leftmargin=1em}
  \item Choose $C_2 \in \LBest$ to maximize $|A(C_2)\backslash A(C_1)|$ over all cuts in $\LBest$.
  \item Choose $C_3$ to be any best answer to that piece $\hat{p} \in \Godd{C_1}{C_2}$ that is closest to $C_2$.
 \end{list}

 \begin{figure}[ht]
  \psfrag{C1}[cc][cc]{$C_1$}
  \psfrag{C2}[cc][cc]{$C_2$}
  \psfrag{ph}[cc][cc]{$\hat{p}$}
  \psfrag{pt}[cc][cc]{$\tilde{p}$}
  \psfrag{oC1C2}[cc][cc]{$\odd{C_1}{C_2}$}
  \psfrag{eC2C1}[cc][cc]{$\even{C_2}{C_1}$}
  \centering
  \includegraphics[width = 0.48\textwidth]{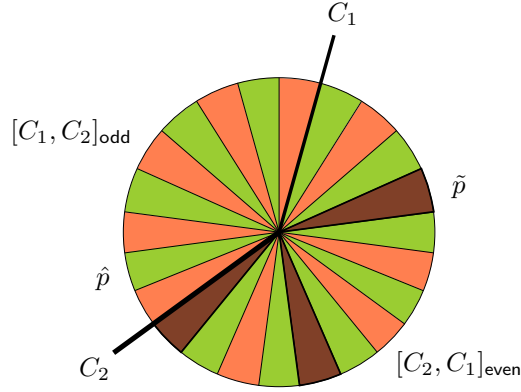}
  \caption{The cuts $C_1$ and $C_2$, the pieces $\hat{p}$ and $\tilde{p}$, and the intervals $\odd{C_1}{C_2}$ and $\even{C_2}{C_1}$ with their canonical colorings. The set $A(C_2)\backslash A(C_1)$ is highlighted.}
  \label{fig:C1-and-C2}
 \end{figure}

 \noindent An example of the situation is depicted in Figure~\ref{fig:C1-and-C2}. We will show that the set $\{C_1,C_2,C_3\}$ satisfies the conditions of the theorem.

 At first, the canonical colorings of $\odd{C_1}{C_1}$ and $\odd{C_2}{C_2}$ agree on $\odd{C_1}{C_2}$ and are reversed on $\even{C_2}{C_1}$. Hence, if $C_2$ is a best answer to a piece $p \in \odd{C_1}{C_2}$, then so is $C_1$. In others words
 \begin{equation}
  A(C_2)\backslash A(C_1) \subseteq \Reven{C_2}{C_1}.\label{eq:A-sets}
 \end{equation}
 Denote by $\tilde{p}$ the last piece of $A(C_2)\backslash A(C_1)$ when going from $C_2$ to $C_1$ through $\even{C_2}{C_1}$. The pieces $\hat{p}$ and $\tilde{p}$ together with the cut $C_2$ divide the pizza into three intervals -- one consisting only of a single piece. This situation is depicted in Figure~\ref{fig:C1-and-C2}.

 \begin{figure}[ht]
  \psfrag{C2}[cc][cc]{$C_2$}
  \psfrag{C3}[cc][cc]{$C_3$}
  \psfrag{ph}[cc][cc]{$\hat{p}$}
  \psfrag{pt}[cc][cc]{$\tilde{p}$}
  \centering
  \includegraphics[width = 1\textwidth]{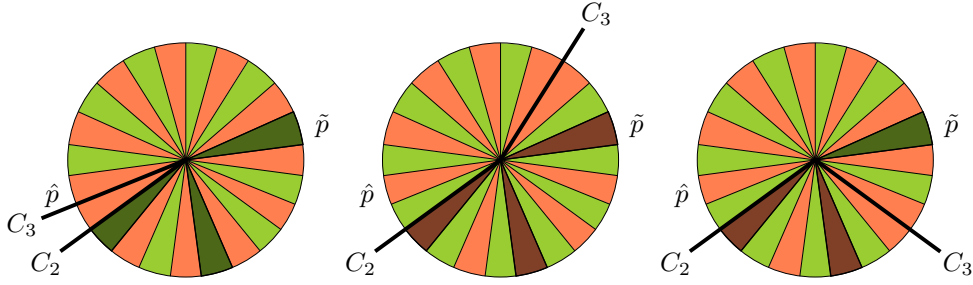}
  \caption{Examples for the theoretically possible positions of $C_3$ together with the canonical coloring of $\odd{C_3}{C_3}$. The set $A(C_2)\backslash A(C_1)$ is highlighted.}
  \label{fig:position-of-C3}
 \end{figure}

 \begin{claim}\label{cla:Lk}
  $C_3$ lies in the interval between $C_2$ and $\tilde{p}$ not containing $\hat{p}$.
 \end{claim}
 \begin{proof}[Proof of Claim]
  Suppose $C_3$ lies in the interval between $\hat{p}$ and $C_2$ not containing $\tilde{p}$ (left-hand case in Figure~\ref{fig:position-of-C3}). Then $C_3$ and $C_2$ are neighboring, since this interval consists of a single piece. Thus the pizza is easy by Lemma~\ref{lem:adjacent-lines} -- a contradiction.

  Suppose $C_3$ lies in the interval between $\hat{p}$ and $\tilde{p}$ not containing $C_2$ (centered case in Figure~\ref{fig:position-of-C3}). Since $C_3$ is a possible answer to $\hat{p}$, we have $\hat{p} \in \even{C_3}{C_2}$ and hence $A(C_2)\backslash A(C_1) \subset \odd{C_3}{C_2}$. More precisely,
  $$A(C_2)\backslash A(C_1) \subseteq \Rodd{C_3}{C_2} \subset \Rodd{C_3}{C_3},$$
  which means that $C_3$ is a possible answer of Bob to all pieces in $A(C_2)\backslash A(C_1)$. Since $C_2$ is a best answer and $C_3$ is a possible answer, $\size{\Rodd{C_2}{C_2}} \leq \size{\Rodd{C_3}{C_3}}$. But since $C_2 \in \LBest$, equality holds and $C_3$ is in $\LBest$ as well. So $C_3$ is a best answer to every piece in $A(C_2)\backslash A(C_1)$ and additionally to $\hat{p} \notin A(C_1)$, contradicting the rule we followed choosing $C_2$.

  We conclude that $C_3$ has to lie according to the right case in Figure~\ref{fig:position-of-C3}.
 \renewcommand{\qedsymbol}{$\bigtriangleup$ \textit{Proof of Claim.}}
 \end{proof}

 \begin{figure}[ht]
  \psfrag{C1}[cc][cc]{$C_1$}
  \psfrag{C2}[cc][cc]{$C_2$}
  \psfrag{C3}[cc][cc]{$C_3$}
  \psfrag{C1C2}[cc][cc]{$\odd{C_1}{C_2}$}
  \psfrag{C1C3}[cc][cc]{$\odd{C_1}{C_3}$}
  \psfrag{C2C3}[cc][cc]{$\odd{C_2}{C_3}$}
  \centering
  \includegraphics[width = .5\textwidth]{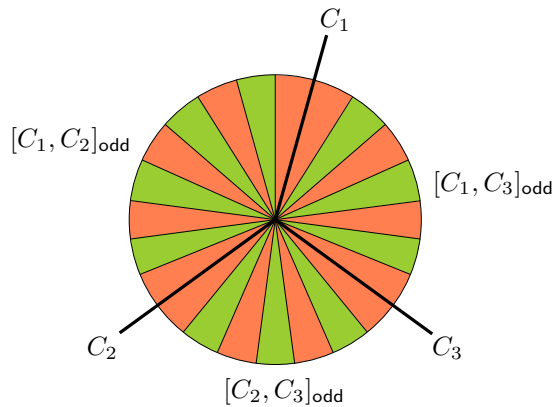}
  \caption{A tripartition of a hard pizza into three odd intervals.}
  \label{fig:tripartition}
 \end{figure}

 By the above claim $C_3$ lies in $\even{C_2}{C_1}$ and since $\hat{p} \in \Rodd{C_3}{C_3}$ the cut $C_3$ splits $\even{C_1}{C_2}$ into two odd intervals, giving a partition into three odd intervals. The result is illustrated in Figure~\ref{fig:tripartition}. Moreover every $C_i$ is a best answer to a green piece in the odd interval opposite to it\footnote{Here we use that $\odd{C_2}{C_3}$ is not just a single red piece, since $C_2$ and $C_3$ are not neighboring by Lemma~\ref{lem:adjacent-lines}.}. With Lemma~\ref{lem:greens} we conclude that each of $\odd{C_1}{C_2}$, $\odd{C_1}{C_3}$, and $\odd{C_2}{C_3}$ has the heavy greens property.
\end{proof}

\section{Best Of Three -- A $3/7$-Strategy}\label{sec:three}

With Theorem~\ref{thm:main-more} we partition a hard pizza by three cuts, each a best answer to a single piece, into three odd intervals, each having the heavy greens property. Based on this tripartition we will now derive a strategy for Alice which guarantees her at least $\frac{3}{7}$ of any pizza.

Let us introduce some abbreviating notation for the total sizes of red and green pieces in each of the three odd intervals. For $i \in \{1,2,3\}$ let $R_i$ be the set of red pieces in the odd interval opposite to $C_i$ and $r_i$ be their total size, e.g., $r_2 = \size{R_2} = \size{\Rodd{C_1}{C_3}}$. Similarly let $G_i$ be the set of \emph{green} pieces in the odd interval opposite to $C_i$ and $g_i$ be their total size, e.g., $g_1 = \size{G_1} = \size{\Godd{C_2}{C_3}}$.

Suppose Alice plays an \fb-strategy associated to some $C \in \{C_1,C_2,C_3\}$, that is she starts with a piece $p$ to which $C$ is a best answer and follows Bob after $p$ is eaten. Doing so Alice gets at least $\size{\Rodd{C}{C}}$ which can be expressed in terms of $r_i$ and $g_i$ as in Table~\ref{tab:follow-bob}.

\begin{table}[ht]
 \begin{tabular}{ccc}
   cut & Alice's outcome & Bob's outcome \\
  \midrule\addlinespace
  $C_1$ & $g_1 + r_2 + r_3$ & $r_1 + g_2 + g_3$\\
  $C_2$ & $r_1 + g_2 + r_3$ & $g_1 + r_2 + g_3$\\
  $C_3$ & $r_1 + r_2 + g_3$ & $g_1 + g_2 + r_3$\\
  \addlinespace
 \end{tabular}
 \caption{Alice's guaranteed outcome achieved by an \fb-strategy associated to $C\in\{C_1,C_2,C_3\}$, respectively. And outcomes of Bob's best reply, respectively.}
 \label{tab:follow-bob}
\end{table}

\noindent As $C_1 \in \LWorst$ we have that Alice's outcome w.r.t. $C_1$ is at most her outcome w.r.t. any \fb-strategy. Similarly as $C_2 \in \LBest$ Alice's outcome w.r.t. $C_2$ is \emph{at least} her outcome w.r.t. any \fb-strategy. In particular we get:
\begin{equation}
 g_1 + r_2 + r_3 \leq r_1 + r_2 + g_3 \leq r_1 + g_2 + r_3 \label{eq:fb-outcomes}
\end{equation}


\begin{rem}
 Since the pizza is hard, each of Alice's outcomes in Table~\ref{tab:follow-bob} is less than $\frac{1}{2}$. This implies $g_i > r_i$ for $i = 1,2,3$, i.e., the green pieces of every odd interval are larger in size than the corresponding red pieces, although they are less.
\end{rem}

\noindent Besides the three \fb-strategies associated to $C_1$, $C_2$, and $C_3$, we will now define three further strategies for Alice, each associated to an odd interval of the tripartition. Note that in each of the \fb-strategies Alice eats the green pieces in one interval and the red pieces in two intervals. The bad case for these outcomes is when the whole pizza lies in the green pieces (this happens in Figures~\ref{fig:follow-Bob-killer} and~\ref{fig:4-9-example}). To improve Alice's guaranteed outcome in general we must provide a way to eat more of the green pieces. To do so, for $i=1,2,3$ let $p_i \in G_i$ be a \emph{middle piece of $G_i$}, that is summing up the sizes of all green pieces from $p_i$ (included) along each direction until hitting a cut in $\{C_1,C_2,C_3\}$ yields at least $\frac{g_i}{2}$.


For $i\in\{1,2,3\}$ the \emph{$i$-th modified-follow-Bob-strategy} denoted as \mfb{i} is defined as follows:

\begin{enumerate}
 \item Alice starts with eating $p_i \in G_i$.
 \item As long as Bob's moves reveal pieces in $G_i$ Alice picks them, i.e., follows Bob.
 \item At the moment Bob's move reveals the first red piece from another of the three odd intervals, Alice makes a single move that does not follow Bob. This means she picks a piece from $R_i$.
 \item Alice follows Bob from then on.
\end{enumerate}


\noindent A modified \fb-strategy contains exactly one move of Alice in which she does not follow Bob. After this particular move some odd interval $\odd{C}{C_j} \subseteq \odd{C_i}{C_j}$ with $C_i \neq C_j \in \{C_1,C_2,C_3\}$ is eaten. So, Alice follows Bob after $\odd{C}{C_j}$ is eaten.

\begin{lem}\label{lem:outside-X}
 Let $C_i, C_j$ be two distinct cuts chosen from $\{C_1,C_2,C_3\}$ and consider $C$ such that $\odd{C}{C_j} \subseteq \odd{C_i}{C_j}$. Then either $C_i$ or $C_j$ is a best answer to $\odd{C}{C_j}$.
\end{lem}

\begin{proof}
 Suppose $\tilde{C} \notin \{C_i,C_j\}$ is a best answer to $\odd{C}{C_j}$. Since $\even{\tilde{C}}{C_j} \subseteq \even{C}{C_j}$ we have $\even{\tilde{C}}{C_i} \subseteq \even{C}{C_j}$ as well\footnote{In particular $\tilde{C}$ does not equal the third cut in $\{C_1,C_2,C_3\}$.}. Hence either $\even{\tilde{C}}{C_i}$ or $\even{\tilde{C}}{C_j}$ is completely contained in an interval of the tripartition (see Figure~\ref{fig:outside-X} for illustration of the three possibilities).

 \begin{figure}[ht]
  \psfrag{C1}[cc][cc]{third in $\{C_1,C_2,C_3\}$}
  \psfrag{C2}[cc][cc]{$C_i$}
  \psfrag{C3}[cc][cc]{$C_j$}
  \psfrag{Ct}[cc][cc]{$\tilde{C}$}
  \psfrag{C}[cc][cc]{$C$}
  \psfrag{C3Ct}[cc][cc]{$\even{C_j}{\tilde{C}}$}
  \psfrag{C2Ct}[cc][cc]{$\even{C_i}{\tilde{C}}$}
  \psfrag{CC3}[cc][cc]{$\odd{C}{C_j}$}
  \centering
  \includegraphics[width = 0.5\textwidth]{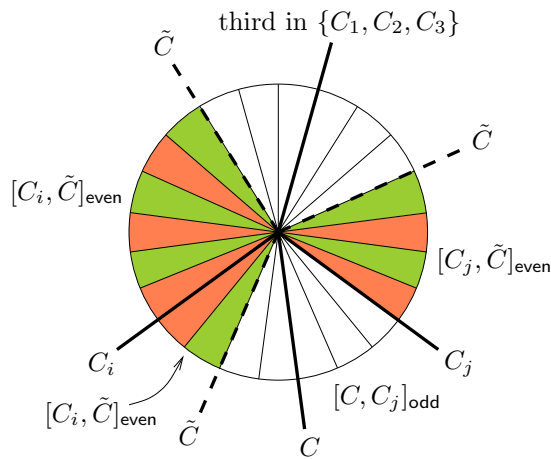}
  \caption{Three cases of the position of $\tilde{C}$ in the proof of Lemma~\ref{lem:outside-X} and the corresponding even interval $\even{C_k}{\tilde{C}}$ with $k\in\{i,j\}$.}
  \label{fig:outside-X}
 \end{figure}

 By Theorem~\ref{thm:main} every interval of the tripartion has the heavy greens property, thus for $k=i$ or $k=j$ we have
 \begin{equation}
  \size{\Geven{C_k}{\tilde{C}}} \geq \size{\Reven{C_k}{\tilde{C}}}. \label{eq:G-geq-R}
 \end{equation}
 \noindent On the other hand $\tilde{C}$ is a best answer and therefore by Lemma~\ref{lem:greens} we have $$\size{\Geven{\tilde{C}}{C_k}} \geq \size{\Reven{\tilde{C}}{C_k}},$$ which is the same as $\size{\Reven{C_k}{\tilde{C}}} \geq \size{\Geven{C_k}{\tilde{C}}}$. Thus equality holds in~\eqref{eq:G-geq-R} and therefore $\size{\Reven{\tilde{C}}{C} \cup \Reven{\tilde{C}}{C_j}} = \size{\Reven{C_k}{C} \cup \Reven{C_k}{C_j}}$. This means that $C_k \in \{C_i,C_j\}$ is a best answer to $\odd{C}{C_j}$, too.
\end{proof}

In an \mfb{}-strategy Alice follows Bob after some $\odd{C}{C_j} \subseteq \odd{C_i}{C_j}$ is eaten. Since by Lemma~\ref{lem:outside-X} either $C_i$ or $C_j$ is a best answer to any $\odd{C}{C_j} \subseteq \odd{C_i}{C_j}$, in a worst case Alice gets either $r_i + g_j$ or $g_i + r_j$ outside of $\odd{C_i}{C_j}$. The inequalities~\eqref{eq:fb-outcomes} imply which possibility has the smaller size and can therefore be assumed to Alice. Together with the definition of the middle piece $p_i$ we then obtain the following guaranteed outcomes for \mfb{}-strategies:

\begin{table}[ht]
\begin{tabular}{ccc}
 \mfb{}-strategy & Alice's outcome\\
 \midrule\addlinespace
 \mfb{1} & $\frac{g_1}{2} + g_2 + r_3$\\
 \mfb{2} & $g_1 + \frac{g_2}{2} + r_3$\\
 \mfb{3} & $g_1 + r_2 + \frac{g_3}{2}$\\
 \addlinespace
\end{tabular}
\caption{Alice's guaranteed outcomes of the three \mfb{}-strategies.}
\label{tab:mfb_X}
\end{table}

%

We have devised, in all, six strategies for Alice: three pure \fb-strategies associated to the cuts $C_1$, $C_2$, and $C_3$ respectively, and three \mfb{}-strategies \mfb{1}, \mfb{2}, and \mfb{3} whose outcomes are bounded from below by Table~\ref{tab:mfb_X}.

Next we show that the best out of these strategies for Alice ensures her at least $\frac{3}{7}$ of the pizza. This can be done by an easy averaging argument.

\begin{thm}\label{thm:3-7}
 Alice can eat at least $\frac{3}{7}$ of any given pizza.
\end{thm}

\begin{proof}
%
 Consider the following strategies for Alice:

 \begin{enumerate}
  \item The \fb-strategy associated to $C_2$, which yields at least $r_1 + g_2 + r_3$ (c.f. Table~\ref{tab:follow-bob});
  \item The \mfb{}-strategy \mfb{1}, which yields at least $\frac{g_1}{2} + r_2 + g_3$ (c.f. Table~\ref{tab:mfb_X});
  \item The \mfb{}-strategy \mfb{3}, which yields at least $g_1 + r_2 + \frac{g_3}{2}$ (c.f. Table~\ref{tab:mfb_X}).
 \end{enumerate}

 Summing up the guaranteed outcomes of \mfb{1}, \mfb{3}, and $\frac{3}{2}$ times the guaranteed outcome of the \fb-strategy associated to the cut $C_2$, we get
 \begin{equation*}
  \frac{3}{2}(r_1 + g_1 + r_2 + g_2 + r_3 + g_3) + \frac{1}{2} r_2 = \frac{3}{2} + \frac{1}{2} r_2.
 \end{equation*}


 Hence the sum of three and a half of Alice's outcomes is at least $\frac{3}{2}$ times the total size of the pizza. Thus, one of the three strategies has to give Alice at least the average value $\frac{3}{2} / \frac{7}{2} = \frac{3}{7}$.
\end{proof}

\begin{rem}
 Restricting Alice to the six above strategies the ratio $\frac{3}{7}$ is tight. Indeed, there is a pizza of which, playing these strategies, Alice eats at most $\frac{3}{7}$. Consider: $g_1=g_2=g_3=2$, $r_1=1$, and $r_2=r_3=0$.
\end{rem}

\section{Best Of Four -- A $4/9$-Strategy}\label{sec:four}

\noindent An \mfb{}-strategy as defined in Section~\ref{sec:three} is composed of a special treatment of one odd interval in the tripartition and following Bob when a certain interval is eaten. In this section we will design strategies for Alice that are particularly devoted to $\odd{C_2}{C_3}$, that is the odd interval in the tripartition that is opposite to $C_1 \in \LWorst$. In order to focus on the essential things, we consider $\odd{C_2}{C_3}$ as a self-contained pizza that arises by cutting off the other two intervals and glueing together the bordering cuts $C_2$ and $C_3$. The resulting pizza we call the \emph{partial pizza} and the glued cut we denote by $C_{2,3}$.



Recall that for a pizza with an odd number of pieces the set $\LWorst$ consists of those cuts $C$ that minimize $\size{\Rodd{C}{C}}$ and hence that are a best answer to every single piece in $\Rodd{C}{C}$. From Lemma~\ref{lem:greens} then follows that $C$ is in $\LWorst$ if and only if $\odd{C}{C}$ has the heavy greens property. Since $\odd{C_2}{C_3}$ has the heavy greens property (Theorem~\ref{thm:main}) we get that the glued cut $C_{2,3}$ of the partial pizza is in $\LWorst$.

\begin{figure}[ht]
 \psfrag{C}[cc][cc]{$C \in \LWorst$}
 \psfrag{Cp}[cc][cc]{$C'$}
 \psfrag{p}[cc][cc]{$p$}
 \psfrag{pp}[cc][cc]{$p'$}
 \psfrag{CCp}[cc][cc]{$\even{C}{C'}$}
 \psfrag{oCCp}[cc][cc]{$\odd{C}{C'}$}
 \centering
 \includegraphics[width = .38\textwidth]{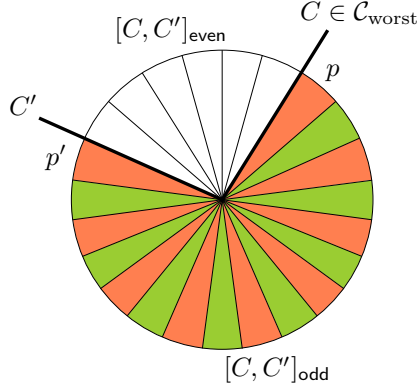}
 \caption{The situation in Lemma~\ref{lem:flip-at-C} and a worst case for the first variant: Alice gets the red pieces $\Rodd{C}{C'}$ and Bob gets the green pieces $\Godd{C}{C'}$.}
 \label{fig:flip-at-C}
\end{figure}

\begin{lem}\label{lem:flip-at-C}
 Consider a pizza with an odd number of pieces (not neccessarily a hard pizza) and an intermediate point in the game at which $\even{C}{C'}$ is eaten for some $C \in \LWorst$. Let $p$ and $p'$ be the pieces that are not in $\even{C}{C'}$ but adjacent to $C$ and $C'$, respectively (see Figure~\ref{fig:flip-at-C}). Then
 \begin{list}{\labelitemi}{\leftmargin=1em}
  \item picking $p'$ and following Bob afterwards
  \end{list}
guarantees Alice at least as much as
 \begin{list}{\labelitemi}{\leftmargin=1em}
  \item picking $p$ and then playing the best she can.
 \end{list}
\end{lem}

\begin{proof}
 Consider the first variant in which Alice picks $p'$ and follows Bob afterwards. Since $C \in \LWorst$ and therefore $\odd{C}{C}$ has the heavy greens property, we get with Lemma~\ref{lem:greens} that $C$ is a best answer to $\even{C}{C'} \cup \{p'\}$. Hence in a worst case of the first variant Alice gets $\Rodd{C}{C'}$ and Bob gets $\Godd{C}{C'}$, which is depicted in Figure~\ref{fig:flip-at-C}.

 Now, the crucial point is that the same distribution can be forced by Bob if Alice plays the second variant, i.e., she picks $p$. To do that, Bob simply \textit{follows Alice} until all the pizza is eaten.

 Therefore by playing the first variant, Alice's guaranteed outcome cannot be worse.
\end{proof}

Lemma~\ref{lem:flip-at-C} enables us to plug valuable strategies for the partial pizza into the strategy for the whole pizza without loosing the guaranteed outcome for Alice. We say that a strategy for the partial pizza is \emph{good} if at the time $C_{2,3}$ is revealed this strategy already tells Alice to follow Bob until the end of the game. Now, a good strategy is said to be \emph{plugged into the whole pizza} if:

\begin{enumerate}
 \item Alice starts in $\odd{C_2}{C_3}$ as if it were just the partial pizza and she pursues the good strategy there.
 \item As long as none of $C_2$ and $C_3$ is revealed, Alice acts according to the good strategy.
 \item In case Bob reveals $C_2$ or $C_3$ (which corresponds to $C_{2,3}$ in the partial pizza), Alice plays the first variant from Lemma~\ref{lem:flip-at-C} instead of the second one.
 \item Whenever Bob picks a piece outside of $\odd{C_2}{C_3}$, Alice follows Bob.
\end{enumerate}

\begin{lem}\label{lem:plugged-into}
 A good strategy that is plugged into the whole pizza ensures Alice at least the guaranteed outcome of the good strategy within the partial pizza plus $r_2 + g_3$.
\end{lem}

\begin{proof}
 As the strategy for the partial pizza is good, Lemma~\ref{lem:flip-at-C} ensures that \emph{inside} $\odd{C_2}{C_3}$ Alice gets at least her guaranteed outcome of this strategy for the partial pizza.

 Alice's outcome \emph{outside} of $\odd{C_2}{C_3}$ can be bounded with Lemma~\ref{lem:outside-X}. Therefore note that Alice follows Bob after some $\odd{C}{C_2} \subseteq \odd{C_2}{C_3}$ or $\odd{C}{C_3} \subseteq \odd{C_2}{C_3}$ is eaten. By Lemma~\ref{lem:outside-X} a best answer is given by either $C_2$ or $C_3$. Thus in a worst case Alice gets either $r_2 + g_3$ or $g_2 + r_3$ outside of $\odd{C_2}{C_3}$. The inequalities~\eqref{eq:fb-outcomes} give $r_2 + g_3 \leq g_2 + r_3$ and hence Alice gets at least $r_2 + g_3$ outside of $\odd{C_2}{C_3}$.
\end{proof}

We conclude with our final theorem.

\begin{thm}
 Alice can eat at least $\frac{4}{9}$ of any given pizza.
\end{thm}

\begin{proof}


 Considering the partial pizza, we distinguish two cases.


 \medskip

 \noindent\emph{Case 1.} The partial pizza is easy.

 By definition there is an \fb-strategy for the partial pizza ensuring Alice at least half of its size. Clearly, every \fb-strategy is good and hence can be plugged into the whole pizza. Consider the following three Alice's strategies for the whole pizza:

 \begin{enumerate}
  \item The \fb-strategy associated to $C_2$, which yields at least $r_1 + g_2 + r_3$ (c.f. Table~\ref{tab:follow-bob});
  \item The \mfb{}-strategy \mfb{2}, which yields at least $g_1 + \frac{g_2}{2} + r_3$ (c.f. Table~\ref{tab:mfb_X});
  \item The \fb-strategy plugged into the whole pizza, which yields at least $\frac{g_1+r_1}{2} + r_2 + g_3$ (c.f. Lemma~\ref{lem:plugged-into}).
 \end{enumerate}

 The claimed $\frac{4}{9}$ of the whole pizza can be proven by calculating an appropriate average out of these strategies: Summing up the guaranteed outcome of $\frac{3}{2}$ times the pure \fb, $2$ times the \fb-strategy plugged into the whole pizza and one outcome of \mfb{2} yields
 \begin{equation*}
  \frac{3}{2}(r_1 + g_2 + r_3) + 2(g_1 + \frac{g_2}{2} + r_3) + (\frac{g_1+r_1}{2} + r_2 + g_3) \geq 2(r_1 + g_1 + r_2 + g_2 + r_3 + g_3) = 2.
 \end{equation*}

 \noindent At least one of the three strategies has to ensure Alice the average value $2 / \frac{9}{2} = \frac{4}{9}$.

 \medskip

 \noindent\emph{Case 2.} The partial pizza is hard.

 By Theorem~\ref{thm:main-more} the partial pizza can be tripartitioned by the cut $C'_1 := C_{2,3} \in \LWorst$ and two further best answers $C'_2$ and $C'_3$ into three disjoint odd intervals each having the heavy greens property. The result is illustrated in Figure~\ref{fig:iterated-tripartition}. We use the natural abbreviating notation for the total sizes of red and green pieces in the three intervals, e.g., $r'_1 = \size{\Rodd{C'_2}{C'_3}}$ and $g'_2 = \size{\Godd{C'_1}{C'_3}}$.


 \begin{figure}[ht]
  \psfrag{C1}[cc][cc]{$C_1$}
  \psfrag{C2}[cc][cc]{$C_2 (=C'_1)$}
  \psfrag{C3}[cc][cc]{$C_3 (=C'_1)$}
  \psfrag{C2p}[cc][cc]{$C'_2$}
  \psfrag{C3p}[cc][cc]{$C'_3$}
  \psfrag{C2pC3p}[cc][cc]{$\odd{C'_2}{C'_3}$}
  \psfrag{C3C2p}[cc][cc]{$\odd{C'_1}{C'_2}$}
  \psfrag{C2C3p}[cc][cc]{$\odd{C'_1}{C'_3}$}
  \psfrag{C1C2}[cc][cc]{$\odd{C_1}{C_2}$}
  \psfrag{C1C3}[cc][cc]{$\odd{C_1}{C_3}$}
  \centering
  \includegraphics[width = 0.58\textwidth]{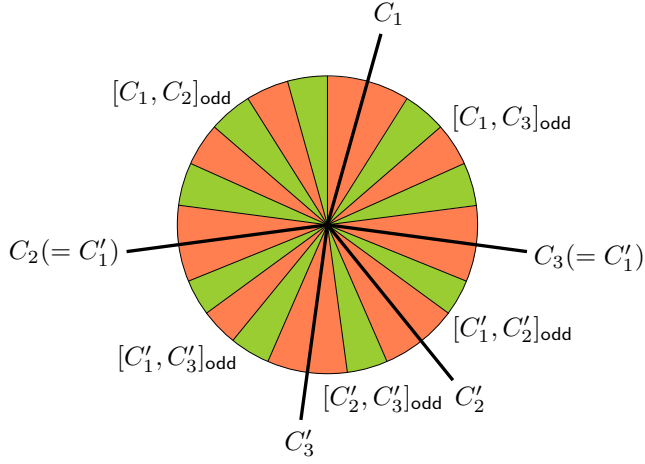}
  \caption{The tripartition of the partial pizza incorporated into the whole pizza yields a partition into five odd intervals each having the heavy greens property. The canonical colorings of the five odd intervals are illustrated.}
  \label{fig:iterated-tripartition}
 \end{figure}

 It is easy to see that the \fb-strategy for the partial pizza that is associated to $C'_2$ as well as the \mfb{}-strategy \mfb{1} for the partial pizza are good and thus can be plugged into the whole pizza. We propose four strategies for Alice for the whole pizza:


 \begin{enumerate}
  \item The \fb-strategy associated to $C_2$, which yields at least $r_1 + g_2 + r_3$ (c.f. Table~\ref{tab:follow-bob});
  \item The \mfb{}-strategy \mfb{2}, which yields at least $g_1 + \frac{g_2}{2} + r_3$ (c.f. Table~\ref{tab:mfb_X});
  \item The \fb-strategy associated to $C'_2$ plugged into the whole pizza, which yields at least $r'_1 + g'_2 + r'_3 + r_2 + g_3$ (c.f. Table~\ref{tab:follow-bob} and Lemma~\ref{lem:plugged-into});
  \item The strategy \mfb{1} for the partial pizza plugged into the whole pizza, which yields at least $\frac{g'_1}{2} + g'_2 + r'_3 + r_2 + g_3$ (c.f. Table~\ref{tab:mfb_X} and Lemma~\ref{lem:plugged-into}).
 \end{enumerate}

 Summing up the guaranteed outcomes of $\frac{3}{2}$ times the first and once the second, the third and the fourth strategy yields
 \begin{equation*}
  \frac{3}{2}(r_1 + g_2 + r_3) + (g_1 + \frac{g_2}{2} + r_3) + (r'_1 + g'_2 + r'_3 + r_2 + g_3) + (\frac{g'_1}{2} + g'_2 + r'_3 + r_2 + g_3).
 \end{equation*}

 \noindent With $r_1 = g'_1 + r_2 + r_3$ and $g_1 = r'_1 + g_2 + g_3$ it follows that the above sum is at least twice the size of the whole pizza. Since we summed up the outcome of $9/2$ strategies, their average value is $2 /\frac{9}{2} = \frac{4}{9}$.
\end{proof}



\noindent \textbf{Final remarks.} If the pieces of the pizza are restricted to be of non-zero minimal size, then Alice can get beyond $\frac{4}{9}$ in any pizza. For this consider the pizza that arises from the given one by shortening every piece by the minimal size of a piece. Afterwards apply our strategy yielding $\frac{4}{9}$ of the smaller pizza. Since Alice eats at least half of the pieces, she definitely eats at least half of the total size that was removed before. Summed up this is strictly more than $\frac{4}{9}$ of the original pizza.

Suppose the pizza is allowed to have pieces of negative size, but the total size of the pizza is positive. Can an outcome for Alice be guaranteed?

A different way of generalizing the problem is to eat other graphs than cycles. One can then ask the eaten or the uneaten part of the graph to remain connected along the course of the game. This question is the topic of \cite{MW-grabbing} and \cite{MW-sharing}.\\

\noindent \textbf{Acknowledgments.} We are indebted to Stefan Felsner, Kamil Kloch and Grzegorz Matecki for many fruitful discussions. Special thanks go to Peter Winkler for stating this catching problem, for being the first reader of the paper and pointing out the extremality of the examples. Thanks to Aaron Dall for his help with the exposition and to Jenny Lettow for designing Figure~\ref{fig:first}.

\bibliographystyle{plain}\bibliography{pizza}

\end{document}